\documentclass[aps,pre,twocolumn,superscriptaddress,longbibliography]{revtex4-2}

\usepackage{graphicx}
\usepackage{amsmath}
\usepackage[percent]{overpic}
\usepackage{hyperref}

\begin{document}


\title{Modeling Optical Polarization Evolution in Myelinated Axon Waveguides with Realistic Imperfections}

\author{Ethan Davies}
\affiliation{Department of Physics and Astronomy, University of Calgary}
\affiliation{Institute for Quantum Science and Technology (IQST), University of Calgary}
\affiliation{Hotchkiss Brain Institute (HBI), University of Calgary}

\author{Rishabh}
\affiliation{Department of Physics and Astronomy, University of Calgary}
\affiliation{Institute for Quantum Science and Technology (IQST), University of Calgary}
\affiliation{Hotchkiss Brain Institute (HBI), University of Calgary}

\author{Christoph Simon}
\affiliation{Department of Physics and Astronomy, University of Calgary}
\affiliation{Institute for Quantum Science and Technology (IQST), University of Calgary}
\affiliation{Hotchkiss Brain Institute (HBI), University of Calgary}

\date{\today}


\begin{abstract}
Biophotonic signaling via axons has been proposed as a potential mode of neural communication, where information might be encoded not only in photon number and wavelength but also in polarization. While there is increasing computational and experimental evidence that axons can guide light, the transmission of polarization is less explored; previous modeling of polarization fidelity in myelinated axons has largely focused on idealized geometries. This study incorporates three structural imperfections characteristic of axons in vivo: variation in myelin thickness, non-circular cross-sectional geometry, and axonal bending, within a model that includes four nodes of Ranvier. We find that variation in myelin thickness alone has minimal impact on fidelity, while non-circular cross-sections show strong mode dependence. Axonal bending has the most significant influence, generating large fluctuations and deep fidelity dips. When all imperfections are combined in a single axon model, the simulations show substantial drops in fidelity, yet certain modes exhibit recovery, with repeated revivals reaching values of around 0.8, which exceeds the revivals observed in the single imperfection cases. Overall, the results indicate that although structural imperfections affect polarization, polarization-information might remain recoverable even in realistic axons. While biophotonic signaling in the brain remains speculative, these results strengthen its physical plausibility. 

\end{abstract}

\maketitle


\section{Introduction}
Neural communication is traditionally understood to occur through electrochemical signaling; however, the physical mechanisms underlying higher-order processes such as consciousness \cite{ref2,ref3,ref4,ref5}, learning \cite{ref6,ref7}, and memory \cite{ref8,ref9} are not completely understood. This uncertainty has motivated interest in whether additional channels of information transfer could coexist with conventional neural signaling. One such proposed channel involves the exchange of biophotons between neurons \cite{ref10,ref11,ref12}. In this context, myelinated axons have been suggested as potential optical waveguides \cite{ref13,ref14,ref15,ref36}, as the myelin sheath exhibits a higher refractive index than both the axon itself and the surrounding extracellular fluid \cite{ref16,ref17}.

Biophotons arise in biological tissues as a consequence of oxidative biochemical processes and span a spectral range from 350 to 1300 nm \cite{ref18}. These ultra-weak photon emissions have been reported across a broad variety of plant and animal cells \cite{ref18,ref19,ref20}, and multiple studies have confirmed their presence within neural tissue and the brain \cite{ref21,ref22,ref23,ref24}. Given that these photons are in principle available to the organism and well suited to serve as signals they could potentially contribute actively to neural information transfer\cite{ref13,ref14,ref18,ref19,ref20,ref13,ref14,ref35}. Supporting this possibility, Opn5-expressing hypothalamic preoptic neurons in mice directly sense violet
light and acutely suppress brown adipose tissue thermogenesis upon illumination \cite{ref25}.

White matter constitutes a large portion of the brain, with myelinated axons serving as the primary nerve fibers that form this tissue \cite{ref27,ref28,ref29}. The myelin sheath electrically insulates the axon, thereby increasing the speed of electrochemical signal transmission between neurons \cite{ref30,ref31,ref32}. Myelination in the central nervous system is carried out by oligodendrocyte glial cells. Glial cells are already known to support light guidance in the vertebrate retina, where Müller cells act as optical waveguides that channel incident visual light toward photoreceptors \cite{ref33,ref34}. This structural arrangement of myelin not only optimizes the speed of electrochemical signals but along with the refractive index of the myelin also provides a good environment for potential optical signaling mechanisms in the brain.

There are several experimental studies that support light propagation along myelinated axons. Early studies reported increased biophotonic activity at the opposite end of isolated nerve roots following optical stimulation \cite{ref92}, and in anatomically connected neural regions following glutamatergic and intracellular optical stimulation \cite{ref90,ref91}. While these studies did not directly image light propagation within axons, they suggested that light-related signals can propagate through neural pathways. More direct experimental evidence was later provided by DePaoli et al., who reported anisotropic light scattering in spinal cord white matter, finding weaker scattering parallel to the local fiber orientation than perpendicular to it, while this directional dependence was absent in gray matter, where myelin is much less abundant and less organized \cite{ref89}. The same study also imaged a longitudinal spinal cord section using transmitted light. A magnified and contrast-enhanced region of this image, shown in Fig. 1, displays bright myelin sheaths surrounding darker axon cores \cite{ref89}. This suggests that the myelin sheath may provide a preferred pathway for light propagation along the axon. More recently, similar anisotropic light propagation was reported in ex vivo human brain white matter, where scattering was likewise found to be weaker parallel to the local fiber orientation than perpendicular to it \cite{ref88}. Together, these observations support the role of the myelin sheath and organized neural fibers in guiding light along nerve pathways.

\begin{figure}[tbp]
    \centering

    \hspace*{0.04\columnwidth} 
    \begin{minipage}{0.45\columnwidth}
        \centering
        \begin{overpic}[width=\linewidth]{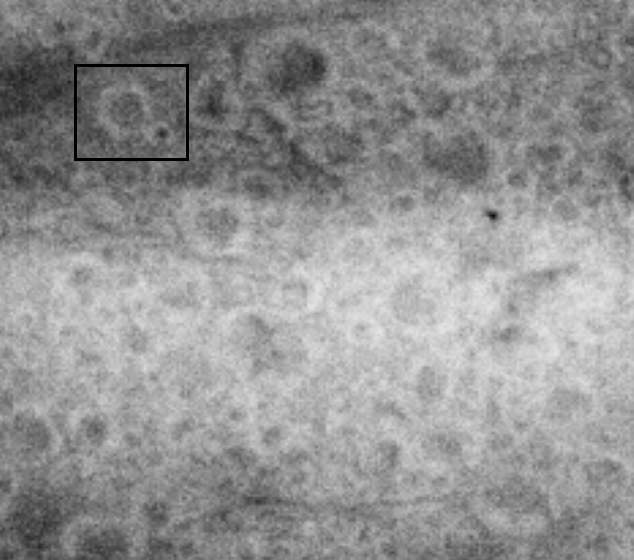}
            \put(-12,84){\small (a)}
        \end{overpic}
    \end{minipage}
    \hfill
    \begin{minipage}{0.40\columnwidth}
        \centering
        \begin{overpic}[width=\linewidth]{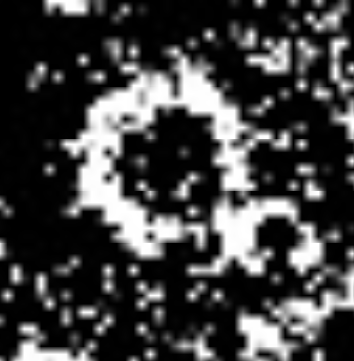}
            \put(-13,95){\small (b)}
        \end{overpic}
    \end{minipage}

    \caption{Images of a $500~\mu\mathrm{m}$-thick longitudinal spinal cord section. (a) Transmission image obtained using transmitted 594-nm light. (b) Magnified and contrast-enhanced image of the boxed region. The bright region corresponds to the myelin sheath, while the darker central region corresponds to the axon core. Panels (a) and (b) are taken directly, without modification, from Figs.~4(c) and 4(b), respectively, of ``Anisotropic light scattering from myelinated axons in the spinal cord'' by DePaoli \textit{et al.} (DOI: 10.1117/1.nph.7.1.015011)~\cite{ref89}. © The Authors of Ref.~\cite{ref89}, licensed under the Creative Commons Attribution 4.0 Unported License.}
    \label{fig:myelin_transmission}
\end{figure}

From a physical perspective, photonic signaling would also offer several potential advantages over electrochemical communication. Although the overall speed of signaling is constrained by emission, detection, and synaptic processes, photonic signals themselves propagate rapidly and, due to their higher energies, are not susceptible to thermal noise at body temperature \cite{ref13,ref14}. Beyond forward signaling, recent theoretical work suggests that axon-guided biophotons could also enable backward transmission of information in neural networks, providing a possible physical mechanism for feedback processes analogous to backpropagation \cite{ref42}.

Biophotonic signals can encode information in various ways, including photon number, frequency, and polarization. Of these, polarization is particularly compelling because it is well-suited for encoding quantum information \cite{ref13,ref14,ref35}, thereby opening up the possibility of quantum communication within the brain\cite{ref13,ref35,ref43}. Due to this, it seems important to assess how well polarization is preserved across axons.

Recent theoretical work has examined biophotonic propagation along myelinated axons \cite{ref15,ref17,ref44,ref45,ref46,ref47,ref48}. In particular, Frede et al.\cite{ref36} presented the first study to consider the evolution of light polarization in myelinated axons, investigating polarization behavior in myelin sheath waveguides containing multiple nodes of Ranvier. The study demonstrated that guided modes can preserve their polarization state across several nodes of Ranvier, providing the first direct evidence that polarization-encoded information could, in principle, survive propagation along a myelinated axon. However, the polarization analysis was carried out using a highly idealized axonal geometry. The axon was represented as a perfectly straight, cylindrical structure with a perfectly circular cross-section and constant myelin thickness. In vivo, axons exhibit bends\cite{ref13,ref37,ref38,ref39}, variations in myelin thickness\cite{ref13,ref40}  and non-circular cross-sectional geometries \cite{ref13,ref39,ref41}. As a result, while this work demonstrates that polarization preservation is possible in idealized myelin-sheath waveguides, the extent to which polarization is preserved under biologically realistic conditions remains unclear. Earlier work by Kumar et al.\cite{ref13} addressed these anatomical imperfections by simulating light guidance in myelinated axons under more biologically realistic conditions. Their model incorporated structural imperfections, including variations in axonal curvature and geometry. While the fundamental principle of optical waveguiding in the myelin sheath was shown to remain valid, these deviations introduced additional scattering, mode distortion, and transmission losses. However, the impact of such biological conditions on the polarization state of guided light was not examined in this study.

Building on these prior studies, the present work extends existing modeling frameworks by incorporating biologically realistic anatomical features, specifically axonal bends, non-circular cross-sectional geometries, and varying myelin sheath thickness, into simulations of biophotonic propagation along myelinated axons, while still including the nodes of Ranvier looked at in previous studies. This model is used to assess whether biophotonic polarization-information can be preserved. under these biologically realistic conditions.

The paper begins by describing the modeling framework and the implementation of the biologically realistic structural imperfections in myelinated axons. The results section then examines the impact of each imperfection individually on polarization preservation, followed by an analysis of their combined effects. The discussion section addresses the implications of these results for the feasibility of polarization-based biophotonic communication in the brain, as well as possible relevance for imaging and medicine.


\section{Model and Methods}

This section will explain each imperfection implemented in this study. After introducing and testing each imperfection in isolation, the final model combines them into a single axon model that simultaneously includes non-circular cross-sections, Ranvier nodes, myelin sheath thickness variation, and axonal bending. An image depicting the axon with all imperfections is shown in Fig. 2.


\begin{figure}[tbp]
\centering
\begin{overpic}[width=1\linewidth]{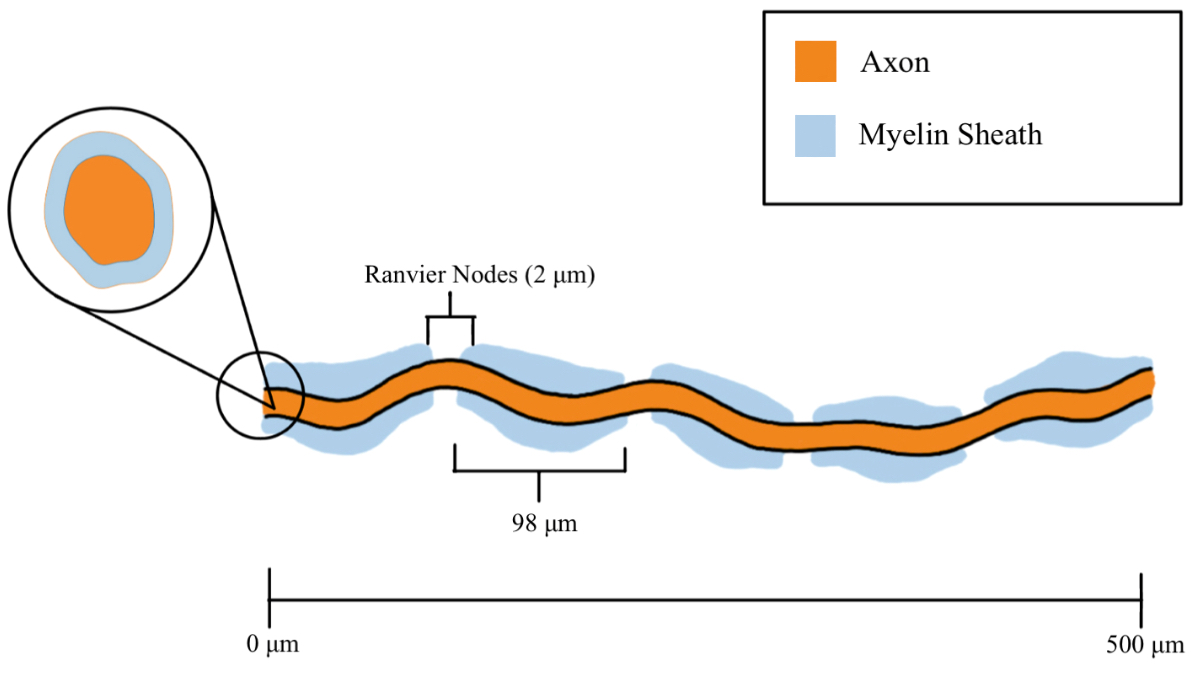}
\end{overpic}
\caption{Artistic diagram of axon incorporating three structural imperfections: irregular
 cross-section, variation in myelin thickness, and axonal bending. The axon is shown in orange,
 and the myelin sheath in blue. The simulation extends from z = 0 µm at the mode source to z
 = 500 µm at the end of the axon. The bends shown here are intentionally exaggerated for visibility.
}
\end{figure}

\subsection{Simulation Region and Monitors}

ANSYS Lumerical FDTD: 3D Electromagnetic Simulator was used to construct the geometries and perform all electromagnetic simulations. The software solves Maxwell’s equations using the finite-difference time-domain (FDTD) method, updating the electric and magnetic fields on a discrete space–time grid within the Yee-cell scheme. This approach enables computation of guided modes and time-domain propagation in myelinated-axon waveguides using nanometer-scale meshes.

The simulation domain spans $500~\mu\mathrm{m}$ in the propagation ($z$) direction, while the transverse ($x$) extent depends on the specific model. The model was limited to this length due to computational resources. A planar mode source is placed at $z=0~\mu\mathrm{m}$, and the myelinated axon is centered in the cross-section and extends through the full domain length so that propagation across multiple Ranvier nodes is included. Perfectly matched layers (PML) are applied on all boundaries to absorb outgoing radiation and represent complete loss at the simulation edges.

To examine polarization evolution, $21$ cross-sectional field monitors were placed along the $500~\mu\mathrm{m}$ propagation length for the geometry containing four Ranvier nodes. Each monitor spans the same $4~\mu\mathrm{m}\times 4~\mu\mathrm{m}$ transverse area as the simulation region and is positioned along $z$ at $25~\mu\mathrm{m}$ spacing, beginning at $z=0~\mu\mathrm{m}$ and continuing to the end of the domain. This spacing is small compared to the $98~\mu\mathrm{m}$ internodal length, providing sufficient resolution to capture field perturbations introduced by the nodes. In Lumerical, monitor locations are snapped to the nearest mesh cell to reduce interpolation; consequently, monitors placed very near node boundaries or close to the simulation end can incur small additional interpolation-related uncertainty.

\subsection{Input light and material optical properties}

In all simulations, the optical wavelength was set to $\lambda=0.4~\mu\mathrm{m}$. This lies within the reported biophotonic spectral window of $0.35~\mu\mathrm{m}$--$1.3~\mu\mathrm{m}$ \cite{ref18}. Due to computational cost a short wavelength was chosen because it is more tightly confined in myelinated axons with thinner sheaths, which in turn allows a smaller simulation region. Additionally, light transmission through the skull is also reduced at shorter wavelengths, with results showing that wavelengths below $\ 550~\mathrm{nm}$ struggle to penetrate the skull \cite{ref87}. This would result in less background light, which could be advantageous for communication based on low-intensity optical signals.

The myelinated axon is modeled using three regions: the axon itself, the myelin sheath, and the surrounding interstitial fluid. Each region is treated as a homogeneous, lossless dielectric with a constant refractive index. The refractive indices used are $n_a=1.38$ for the axon \cite{ref74}, $n_m=1.44$ for the myelin sheath \cite{ref16}, and $n_{\mathrm{int}}=1.34$ for the interstitial fluid \cite{ref75}. This index contrast, with $n_m$ having the highest index, provides the basic physical mechanism that enables optical waveguiding in the myelin sheath.

It is assumed that absorption by the myelin sheath is negligible at the chosen wavelength and, more broadly, across the biophotonic range over centimeter-scale distances\cite{ref13}. Myelin consists mainly of lipids, proteins, and water, each of which has weak absorption in this band. Mammalian fat (a close proxy for myelin lipids) has an absorption coefficient below $0.01~\mathrm{mm^{-1}}$ across the biophotonic spectrum \cite{ref76}, and water absorption is similarly low. Most proteins, including those found in myelin, show minimal absorption for wavelengths above $\sim 0.34~\mu\mathrm{m}$ \cite{ref77}. This assumption is also supported by measured optical properties of brain tissue: in white matter, the absorption coefficient decreases from about $0.3~\mathrm{mm^{-1}}$ to $0.07~\mathrm{mm^{-1}}$ over $0.4~\mu\mathrm{m}$--$1.1~\mu\mathrm{m}$ \cite{ref78}, and grey matter exhibits comparable absorption, suggesting myelin does not strongly increase it. Finally, experimental studies indicate that attenuation in brain tissue is dominated primarily by scattering rather than absorption \cite{ref79}.

\subsection{Varying myelin sheath thickness}

Ranvier nodes are short, unmyelinated segments of axon that interrupt the myelinated internodal regions at regular intervals and therefore play an essential role in assessing waveguiding in the myelin sheath. Node lengths are typically $1$–$2~\mu\mathrm{m}$ \cite{ref82,ref83}; all models in this study have a set node length of $2~\mu\mathrm{m}$ and implement each node as an abrupt termination of the myelin sheath, so that the nodal cross-section contains only the axon, as shown in Fig. 2. To limit the computational cost, the models do not include paranodal tapering at the internode--node boundaries. Each model in this study has a total propagation length of $L=500~\mu\mathrm{m}$ each with a node-to-node spacing of $100~\mu\mathrm{m}$, which yields internodes of length $98~\mu\mathrm{m}$ between adjacent $2~\mu\mathrm{m}$ nodes. Accordingly, the structure contains four Ranvier nodes located at $z=100$, $200$, $300$, and $400~\mu\mathrm{m}$.

Within each internode, certain models in this study incorporate longitudinal myelin-thickness variation through a smoothly varying $g$-ratio, defined as $g(z)=r_a/(r_a+t(z))$, where $t(z)$ is the local myelin thickness. Studies report a mean $g$-ratio of 0.69 for corpus callosum axons in human subjects, consistent with analytically predicted optimal values in the range $0.6$--$0.77$ \cite{ref84}. Motivated by observations that sheath thickness varies less along the internode in mature rat sheaths (often peaking near the midpoint) \cite{ref40}, the g-ratio is constrained to $g\in[0.60,0.80]$ and implemented by prescribing a smooth bell-shaped thickness profile $t(z)$ with thinner myelin at the internode edges and thicker myelin near the center, while keeping the axon boundary fixed. Each internode is discretized into $n_{\rm eval}=1600$ axial slices, and uses a smooth bell-shaped $g(z)$ profile whose edge and midpoint values are drawn from a random subrange within $g\in[0.60,0.80]$; this is converted to thickness via $t(z)=r_a\left(1/g(z)-1\right)$ and applied slice-by-slice by scaling only the outer myelin boundary while keeping the axon radius fixed. For only the first internode, the edge value is intentionally anchored at $g_{\rm edge}=0.70$ to ensure consistent modes at launch for each simulation The resulting g-ratio (thickness) profile used in simulations is shown in Fig. 3.

\begin{figure}[tbp]
\centering

\begin{minipage}{1\linewidth}
\begin{overpic}[width=\linewidth]{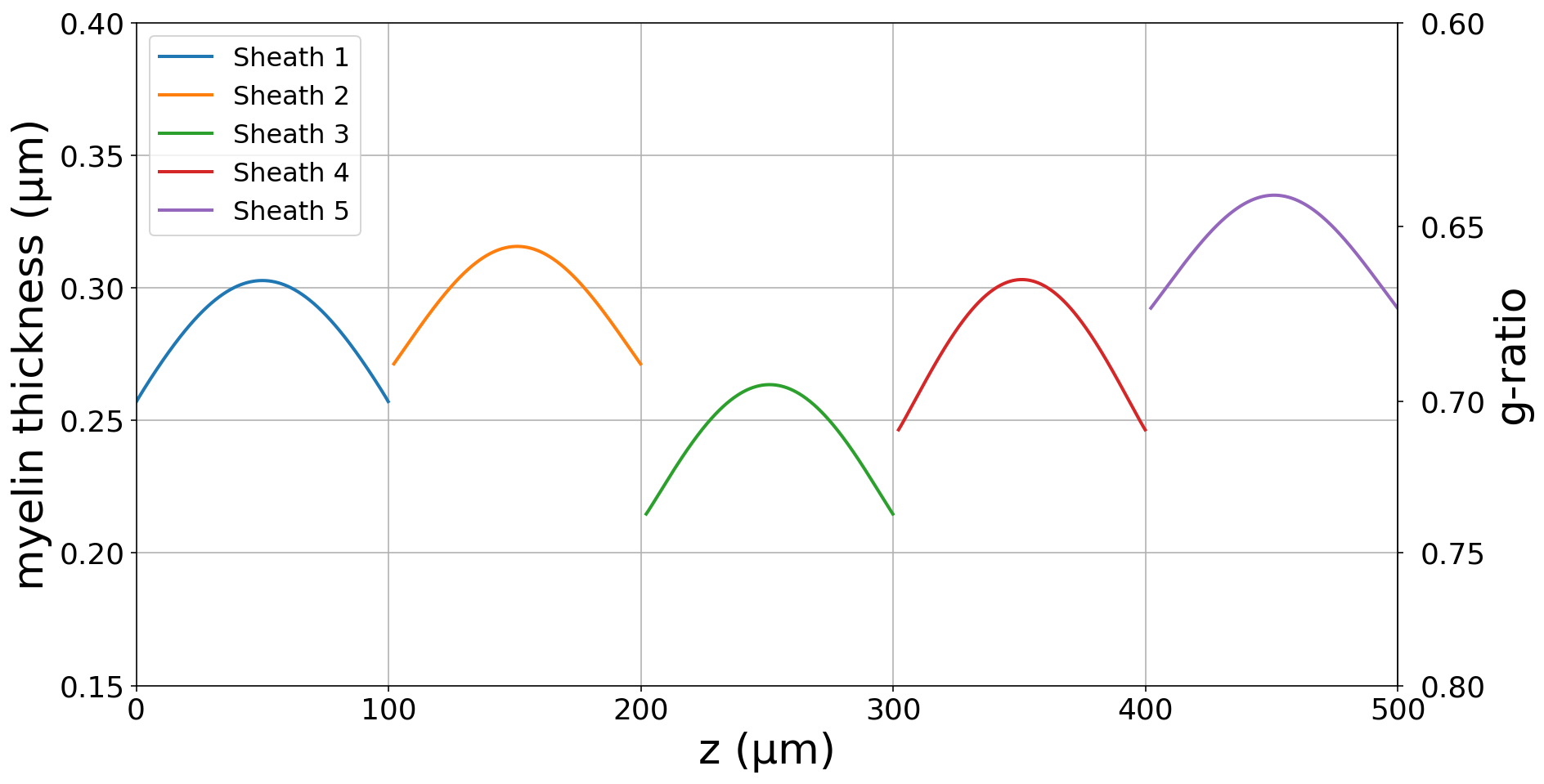}
\end{overpic}
\end{minipage}

\caption{Myelin thickness variation used in simulations.}
\end{figure}

\subsection{Non-circular cross-sectional area}

Studies indicate that axons are more accurately described as slightly elliptical rather than perfectly circular \cite{ref80}. To move beyond an idealized circular nerve-fiber cross-section, the axon is modeled as a closed polygon that captures both ellipticity and small, biologically motivated deviations from a perfect ellipse. We set the axon radius to $r_a=0.6~\mu\mathrm{m}$ and impose ellipticity by choosing semiaxes $a_{\rm in}=r_a/1.15$ and $b_{\rm in}=1.15r_a$, this value of $r_a$ is selected to balance computational cost while remaining consistent with reported cortical white-matter axon diameters spanning approximately $0.2~\mu\mathrm{m}$--$10~\mu\mathrm{m}$ \cite{ref81}. To mimic natural geometric irregularity, the boundary is formed by adding a small band-limited Fourier perturbation to the radius as a function of angle using low-order $\sin$ and $\cos$ modes with randomly drawn coefficients, and discretizing the resulting smooth contour with $n=720$ points. The resulting polygon defines the axon boundary, and a corresponding baseline sheath polygon is then scaled to generate the local outer myelin boundary used throughout the model. Cross-sections used in this work are shown in Fig. 4.

\begin{figure}[tbp]
\centering

\begin{minipage}{0.46\linewidth}
\begin{overpic}[width=\linewidth]{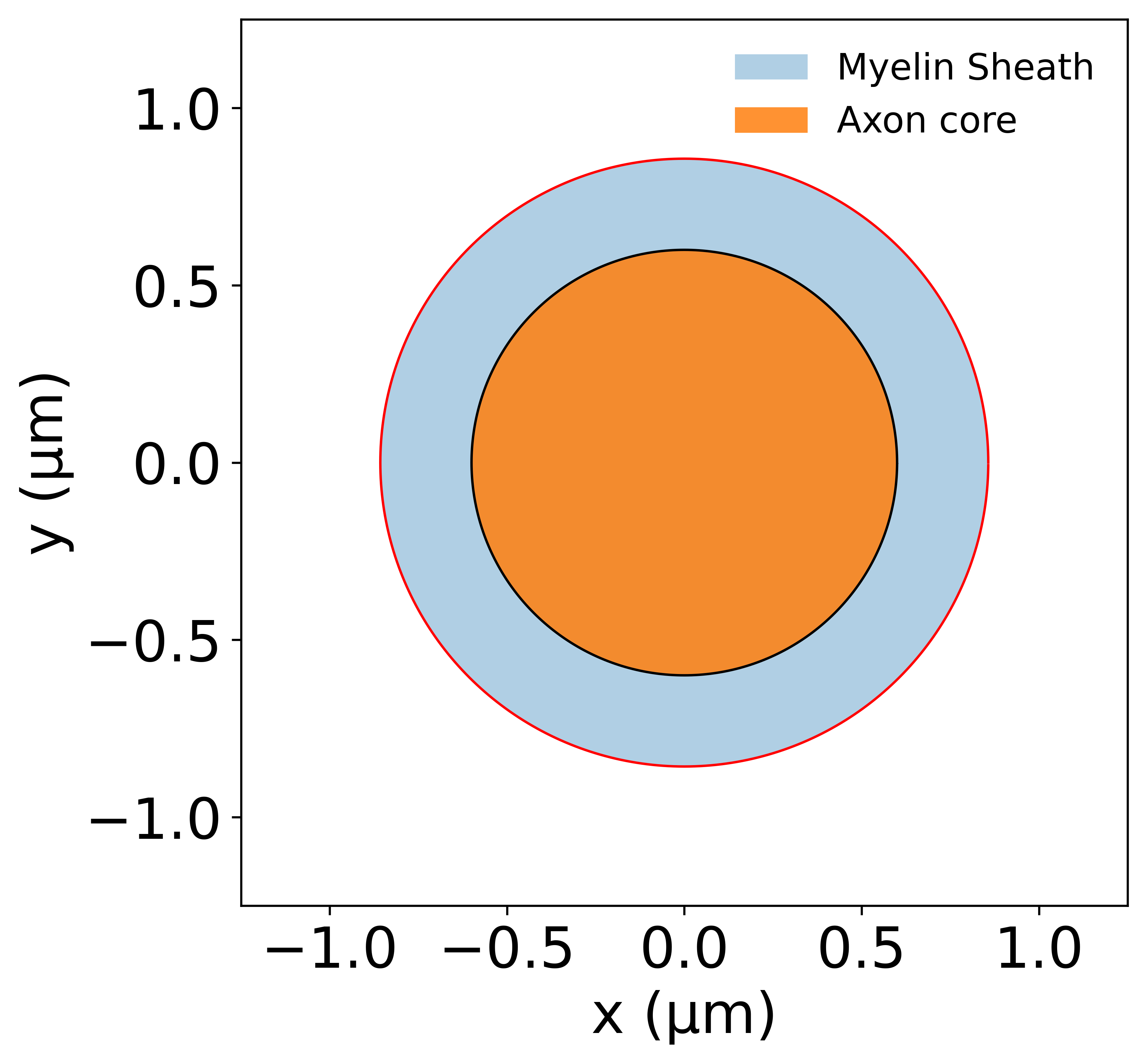}
    \put(-4,85){\small (a)}
\end{overpic}
\end{minipage}
\hspace{0.04\linewidth}
\begin{minipage}{0.46\linewidth}
\begin{overpic}[width=\linewidth]{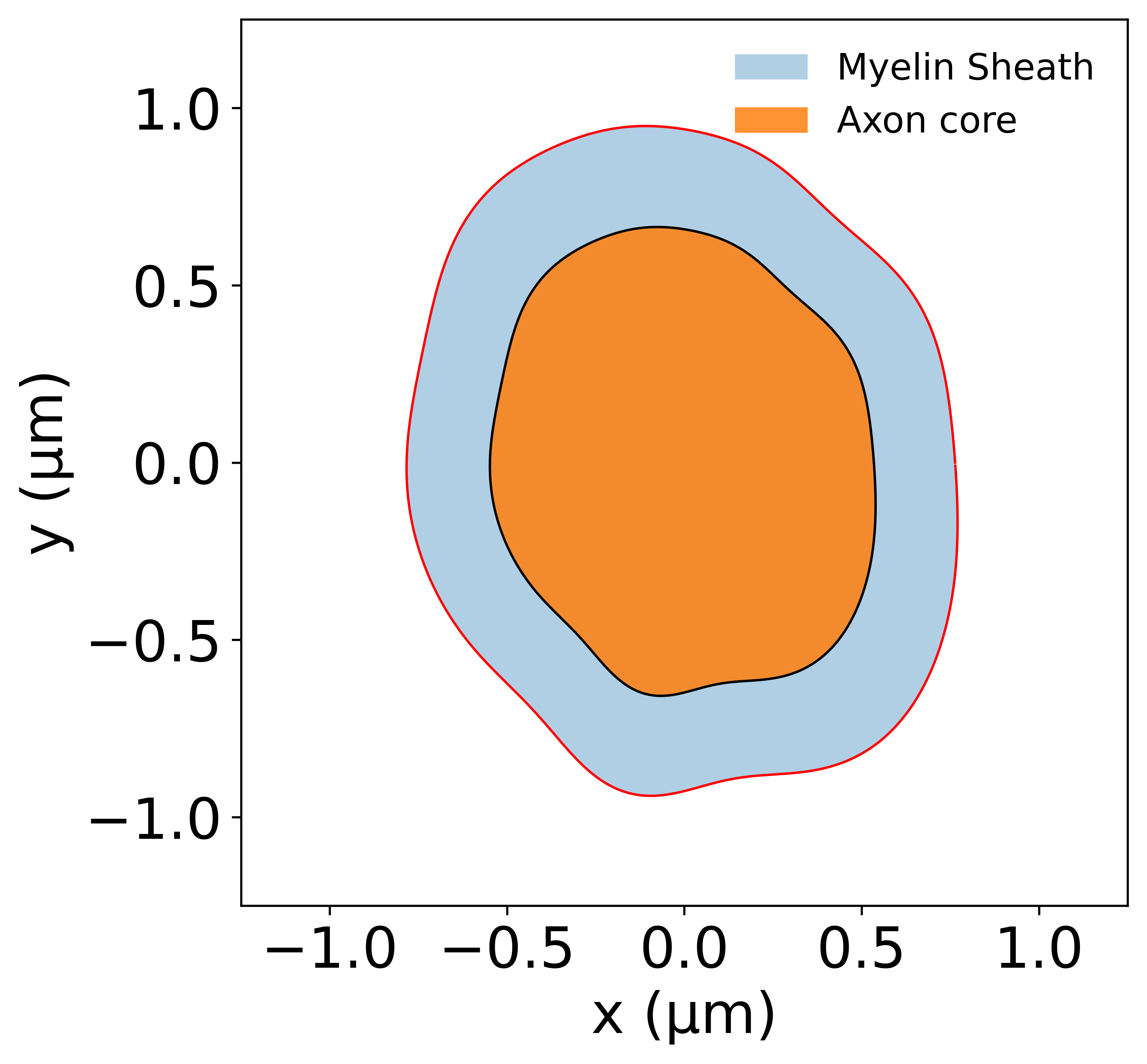}
    \put(-4,85){\small (b)}
\end{overpic}
\end{minipage}

\caption{Cross-sectional geometries used in simulations.  
(a) Perfect circle axon cross-section.  
(b) Irregular, non-circular cross-section.
The axon core (orange) is outlined in black (axon boundary), and the myelin sheath (blue) is outlined in red (outer myelin boundary).
}
\end{figure}


\subsection{Bending}

A study of white-matter axons in monkey brains reported that corpus callosum axons exhibit both smaller local bends and larger-scale curvature with an overall median tortuosity of $\tau=1.01$, one of the lowest values documented \cite{ref85}. To model bending, a smoothly varying axonal centerline is generated and each axon/myelin cross-section is translated along this trajectory, thereby producing a curved waveguide geometry. The centerline includes a macro-bend with a maximum lateral shift set to keep the geometry within the computational limit ( $\Delta_{\max}=11~\mu\mathrm{m}$), and additional smaller bends generated as smoothed random fluctuations along the axon length. The overall bend strength is tuned such that the resulting path satisfies $\tau=L_{\rm path}/L=1.01$. The centerline x-offset profile corresponding to $\tau=L_{\rm path}/L=1.01$ is shown in Fig. 5.

\begin{figure}[tbp]
\centering

\begin{minipage}{1.02\linewidth}
\begin{overpic}[width=\linewidth]{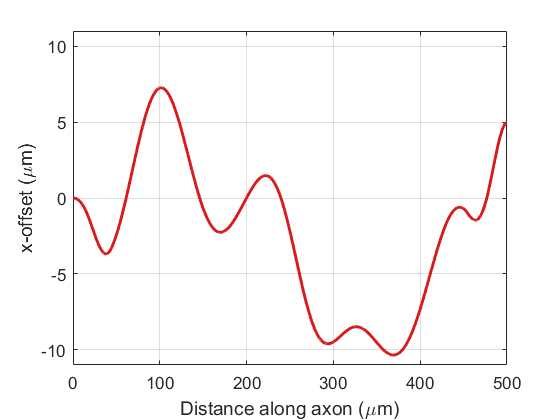}
\end{overpic}
\end{minipage}

\caption{Centre-line x-offset profile along the axon for a tortuosity of 1.01. The plotted bends appear visually large because of the axis scaling.}
\end{figure}


\section{Results}

In this section, simulation data are presented for five axon models that isolate the impact of individual imperfections, followed by a composite model that combines all imperfections. At each of the 21 monitors along the axon, the electric field profile is recorded and used to quantify polarization evolution via the normalized overlap with the input field profile. The polarization fidelity $F(z)$, as a function of longitudinal position $z$, is defined as follows:

\begin{equation}
\resizebox{0.8\columnwidth}{!}{$
F(z) =
\frac{\left|\displaystyle\iint \vec{E}(x,y,0)\cdot\vec{E}(x,y,z)\,dx\,dy\right|}
{\sqrt{\displaystyle\iint \left|\vec{E}(x,y,0)\right|^2\,dx\,dy\;\displaystyle\iint \left|\vec{E}(x,y,z)\right|^2\,dx\,dy}}
$}
\label{eq:fidelity}
\end{equation}

Each figure corresponding to an isolated imperfection includes results for two injected input modes. For each mode, the output intensity distribution and the transverse vector-field profile are shown, together with a plot of the polarization transmission fidelity F(z) comparing the two inputs. The same fidelity plot format is used for the model that combines all imperfections.

For the circular and non-circular cross-section models, the two injected modes were selected to allow for a structural comparison. Specifically, modes were chosen where the mode structure was as similar as possible between the circular and non-circular geometries. Here, the mode structure is the intensity profile and polarization field. Though it should be noted that meaningful comparison between circular and non-circular geometries is difficult due to how drastically geometry impacts mode structure. Modes used for the perfectly circular cross-section are given by Fig. 6 (a-d), and modes used for the non-circular cross-section are given by Fig. 6 (e-h). The non-circular geometry modes were chosen because they best illustrated mode dependence: one mode exhibits relatively strong polarization preservation, while the other does not. The non-circular cross-section breaks rotational symmetry, so the guided modes are no longer evenly distributed around the sheath and instead concentrate into localized lobes in the myelin.

\begin{figure}[tbp]
\centering


\begin{minipage}{0.45\linewidth}
\begin{overpic}[width=\linewidth]{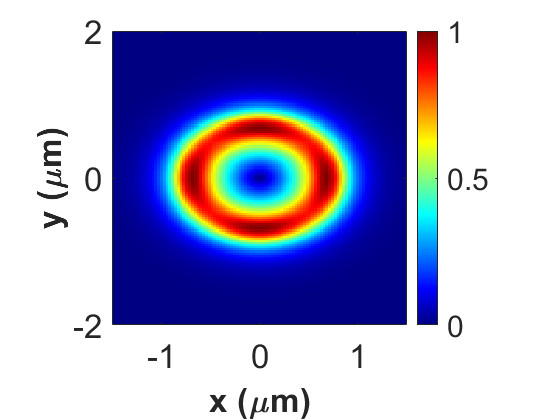}
    \put(0,67){\small (a)}
\end{overpic}
\end{minipage}
\hspace{0.02\linewidth}
\begin{minipage}{0.45\linewidth}
\begin{overpic}[width=\linewidth]{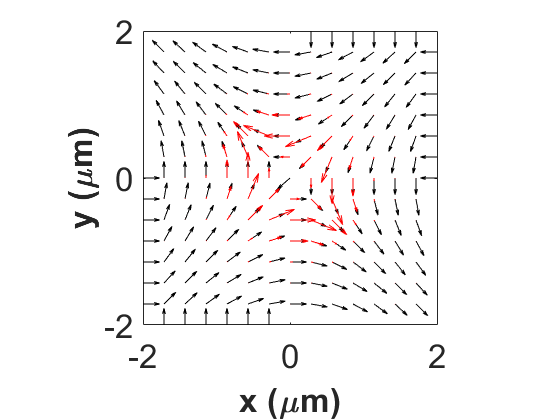}
    \put(0,67){\small (b)}
\end{overpic}
\end{minipage}

\vspace{-0.1em}

\begin{minipage}{0.45\linewidth}
\begin{overpic}[width=\linewidth]{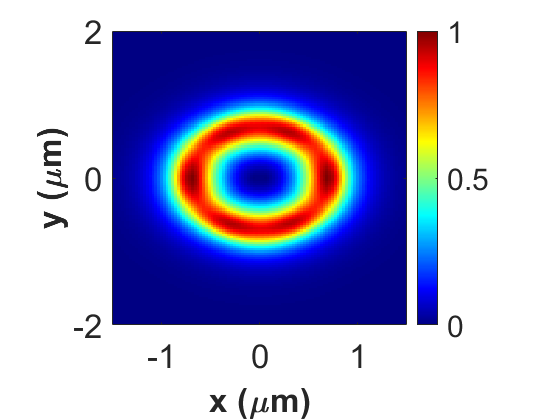}
    \put(0,67){\small (c)}
\end{overpic}
\end{minipage}
\hspace{0.02\linewidth}
\begin{minipage}{0.45\linewidth}
\begin{overpic}[width=\linewidth]{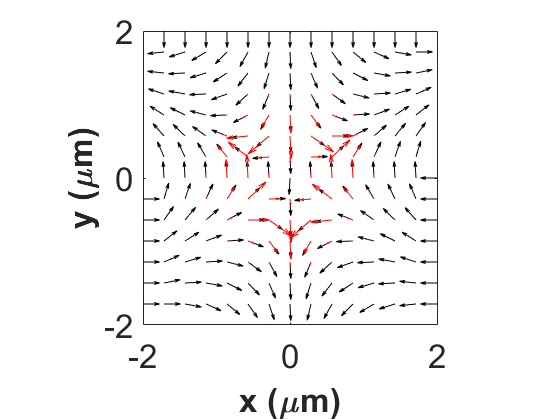}
    \put(0,67){\small (d)}
\end{overpic}
\end{minipage}

\vspace{0.4em}


\begin{minipage}{0.45\linewidth}
\begin{overpic}[width=\linewidth]{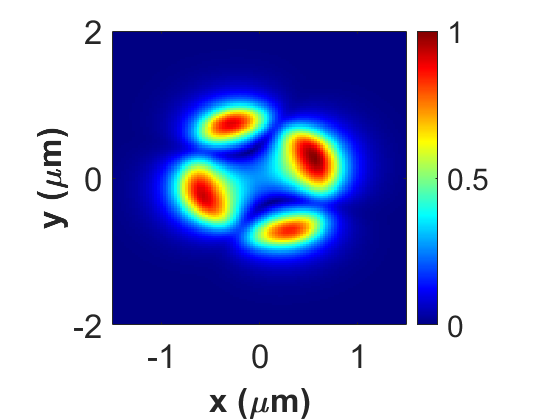}
    \put(0,67){\small (e)}
\end{overpic}
\end{minipage}
\hspace{0.02\linewidth}
\begin{minipage}{0.45\linewidth}
\begin{overpic}[width=\linewidth]{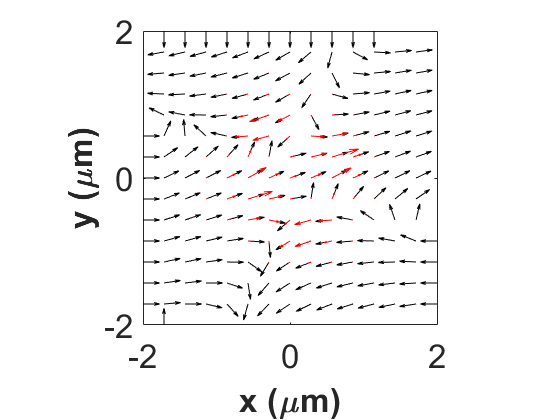}
    \put(0,67){\small (f)}
\end{overpic}
\end{minipage}

\vspace{-0.1em}

\begin{minipage}{0.45\linewidth}
\begin{overpic}[width=\linewidth]{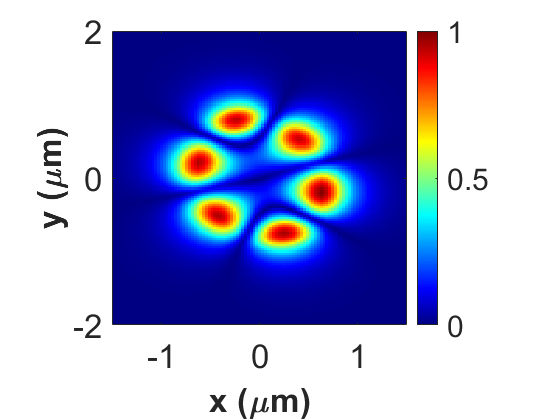}
    \put(0,67){\small (g)}
\end{overpic}
\end{minipage}
\hspace{0.02\linewidth}
\begin{minipage}{0.45\linewidth}
\begin{overpic}[width=\linewidth]{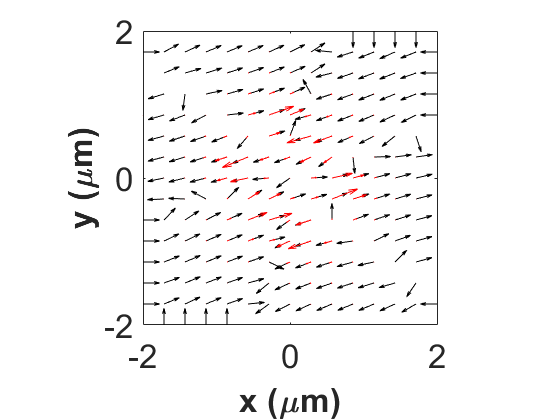}
    \put(0,67){\small (h)}
\end{overpic}
\end{minipage}

\caption{Injected modes used for the circular and non-circular cross-section models. (a,c) Intensity profiles of Mode 1 and Mode 2 for the perfectly circular cross-section model. (b,d) Corresponding polarization vector fields. (e,g) Intensity profiles of Mode 1 and Mode 2 for the non-circular cross-section model. (f,h) Corresponding polarization vector fields.}
\label{fig:circular_noncircular_modes}
\end{figure}

Fig. 7 shows results for the control axon model with no imperfections (straight, circular cross-section, and uniform myelin thickness). Both injected modes maintain high fidelity over the full propagation length. The radius of the axon is set to $r_a=0.6~\mu\mathrm{m}$ with a constant g-ratio of 0.7. The modes for the perfectly circular cross-section shown in Fig. 6 (a-d) are used for this model.
\begin{figure}[tbp]
\centering
\includegraphics[width=0.95\linewidth]{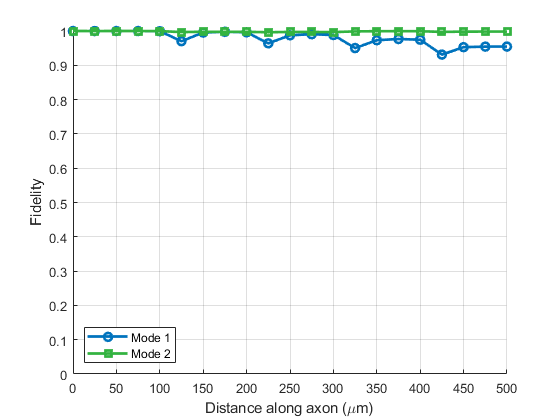}
\caption{Polarization fidelity for control axon for the two injected modes along the axon.}
\label{fig:control_fidelity}
\end{figure}

Fig. 8 shows results for the model where myelin-thickness variation is the only imperfection. This model remains straight with a perfectly circular cross-section, but the myelin thickness varies as shown in Fig. 2. The resulting fidelity results are similar to those of the control model for both injected modes with the difference of small fidelity dips near the end of the axon. The modes for the perfectly circular cross-section shown in Fig. 6 (a-d) are used for this model.

\begin{figure}[tbp]
\centering
\includegraphics[width=0.95\linewidth]{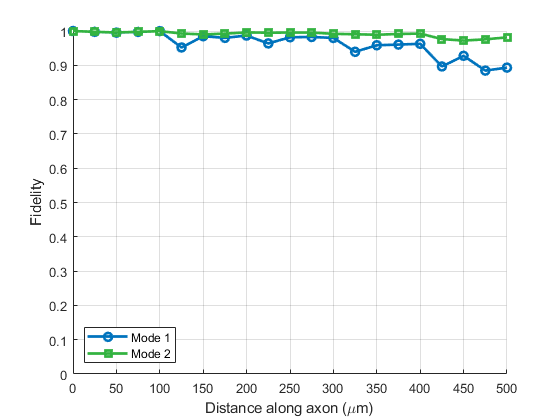}
\caption{Polarization fidelity for axon with varying myelin thickness for the two injected modes along the axon.}
\label{fig:thickness_fidelity}
\end{figure}

Fig. 9 shows results for the model with a non-circular cross-section as the only imperfection. The axon remains straight with uniform myelin thickness, but the cross-section is non-circular as shown in Fig. 3. In this case, the fidelity shows mode dependence, with mode 2 maintaining higher polarization fidelity compared to mode 1. The modes for the non-circular cross-section shown in Fig. 6 (e-h) are used for this model.


\begin{figure}[tbp]
\centering
\includegraphics[width=0.95\linewidth]{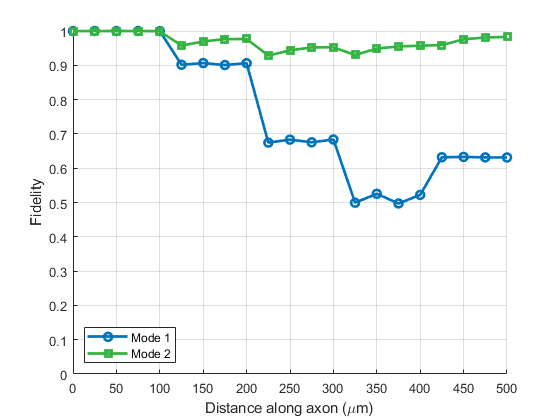}
\caption{Polarization fidelity for the axon with a non-circular cross-section for the two injected modes along the axon.}
\label{fig:noncircular_fidelity}
\end{figure}

Fig. 10 shows results for the model with axonal bending as the only imperfection. The cross-section remains perfectly circular and the myelin thickness is constant, but the axon centerline is bent as shown in Fig. 5. In this case, both injected modes exhibit pronounced drops and fluctuations in fidelity along the propagation length. The modes for the perfectly circular cross-section shown in Fig. 6 (a-d) are used for this model.


\begin{figure}[tbp]
\centering
\includegraphics[width=0.95\linewidth]{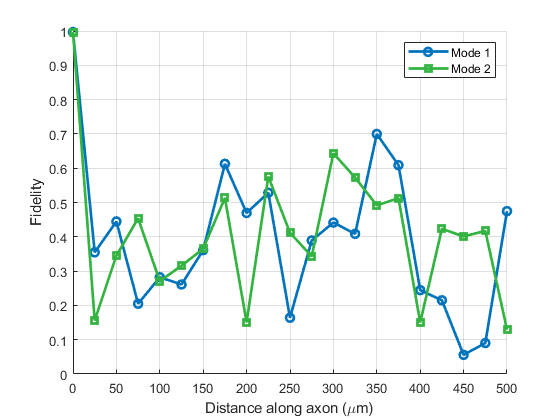}
\caption{Polarization fidelity for the bent axon with tortuosity of 1.01 for the two injected modes along the axon.}
\label{fig:bending_fidelity}
\end{figure}

Fig. 11 shows results for the model that combines all previously considered imperfections. The axon is bent, has a non-circular cross-section, and includes spatially varying myelin thickness. Fidelity exhibits large drops and pronounced fluctuations for both injected modes; however, Mode 2 shows partial revivals to higher fidelity values multiple times across the propagation distance. Interestingly, these revivals are greater than the revivals seen in the bending in isolation case. The modes for the non-circular cross-section shown in Fig. 6 (e-h) are used for this model.

\begin{figure}[tbp]
\centering
\includegraphics[width=0.95\linewidth]{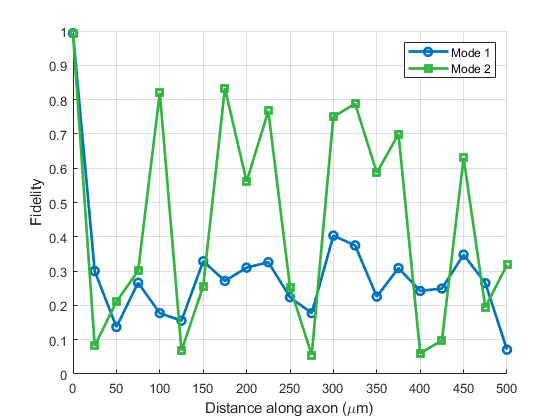}
\caption{Polarization fidelity for the axon containing all imperfections for the two injected modes along the axon.}
\label{fig:all_imperfections_fidelity}
\end{figure}


\section{Discussion}
The present work demonstrates that incorporating realistic anatomical features, including variation in myelin thickness, non-circular cross-sections, and axonal bending, substantially changes the polarization dynamics of guided light in myelinated axons. Across all simulations, these anatomical imperfections introduce fluctuations and pronounced dips in polarization fidelity. However, in the axon model that contains all considered anatomical imperfections, while the polarization fidelity shows large fluctuations, it also shows consistent revivals to higher values, repeatedly reaching levels around 0.8 for certain modes. This behavior suggests that even in the presence of multiple realistic imperfections, hypothetical polarization-encoded signals may remain at least partially recoverable. Importantly, these results highlight strong mode dependence: some guided modes preserve polarization fidelity substantially better than others. Some modes might couple only weakly (or mainly to just one or two other modes), so the light doesn’t spread into many different modes. If that’s the case, the energy can cycle back into the original component, giving those revivals seen for mode 2 of Fig. 10. In contrast, stronger coupling into many modes would spread the energy out and make revivals less likely. Other supported mode structures not explored here may exhibit even greater fidelity preservation under the same anatomical imperfections.

Although this model is an important step in the pursuit of a realistic axon model, several features could be added to more accurately represent realistic biophoton propagation in myelinated axons. First, light detection in the model is implemented using planar monitors that sample entire cross-sections. However, biological detectors would likely be localized in nature \cite{ref36}. Proposed light-sensitive structures include centrosomes\cite{ref54}, chromophores\cite{ref55} and opsins\cite{ref25}. Second, only a single guided mode is launched. This neglects the discrete, point-like nature of biophoton emission and does not address how photons produced inside the axon actually couple into the guided modes of the myelin sheath. In neurons, candidate sources of biophotons include mitochondria \cite{ref49,ref50,ref51}, via mitochondrial respiration, and liposomes\cite{ref52}, via lipid oxidation. Mitochondria are especially interesting in this context because recent studies suggest that they can encode information about neuron firing frequency\cite{ref53}, raising the possibility that biophoton production could be linked to action potential signaling. Together, these considerations indicate that the current framework does not yet capture the inherently localized nature of both emission and detection in vivo.

There are still more features that future extensions could incorporate to more accurately simulate in vivo conditions, such as increasing the simulated axon length beyond 500 µm, as myelinated axons are longer in biological tissue. For instance, callosal fibres linking primary somatosensory cortices in macaque monkeys can reach lengths of roughly 2 cm \cite{ref56}. While this would improve biological realism, it would also increase the computational cost of the simulations. Future work could also explore a broader range of axon radii and optical wavelengths. Other additions could include introducing more structural heterogeneity by allowing the lengths of myelin sheaths and nodes of Ranvier to vary. The model could also incorporate paranodal regions, which are transition zones at each Ranvier node where the myelin sheath gradually thins and ends \cite{ref13}, as well as the positive radial birefringence of the myelin sheath \cite{ref57,ref58}. Incorporating additional sources of optical scattering would provide a further level of realism. These could include intracellular scatterers such as mitochondria\cite{ref72}, as well as refractive-index heterogeneity within the myelin sheath\cite{ref73}. The model treats each region as homogeneous with a constant refractive index, but real tissue is not perfectly uniform, so additional index variations could increase scattering.

Experimental studies indicate that neural tissue is broadly light sensitive\cite{ref25,ref63,ref64,ref65,ref66}, which motivates examining whether such photonic activity could support information transfer in the brain. The central question in this work is whether polarization can remain sufficiently stable as light travels through the myelin sheath to act as a reliable information channel between neurons. This would enable photonic signals to carry classical information in neural tissue. Furthermore, biophotons could also, at least in principle, participate in communication that involves quantum information, since polarization naturally supports the encoding of qubits in addition to classical bits. Recent studies have explored whether quantum effects could play functional roles in biology, including in the brain \cite{ref59}. There is potential that quantum information transmitted through biophotons may be able to maintain entangled systems of spins within neurons \cite{ref35}. From this viewpoint, quantum information processing in neural tissue \cite{ref60} could offer advantages similar to those demonstrated in quantum computing \cite{ref61}. This advantage has been suggested as one way the brain might meet high computational demands while operating under strict energy restrictions\cite{ref36}. Some studies further speculate that quantum entanglement could explain the complex nature of consciousnesses \cite{ref62,ref35}. However, establishing the feasibility of quantum communication would also require demonstrating that superpositions of different polarization states can be preserved during propagation, which was not examined in the present work and remains an important direction for future study.

Motivations for characterizing polarization evolution in axons extend beyond the possibility of biophotonic communication. The optical properties of the myelin sheath are also relevant in the context of neuroimaging. Polarized light forms the basis of 3D polarized light imaging (3D-PLI), which enables microscopic imaging of nerve fibers \cite{ref67,ref68}. Diseases associated with myelin, such as multiple sclerosis, could be further understood by investigating the optical properties of the myelin sheath, which could be significant for medical advancement \cite{ref71}. 

We hope that our modeling results will inspire experimental efforts to characterize light propagation and polarization preservation in axons, which will lead to further progress in our understanding of the potential for classical and quantum biophotonic communication, as well as having implications for medicine.

\section*{Data Availability Statement}

The data and code supporting the findings of this study are available in Ref~\cite{davies_myelin_data_2026}.

\section*{acknowledgments}
This work was supported by the Natural Sciences and Engineering Research Council (NSERC) of Canada via its Discovery Grant program and the Alliance Quantum Consortia grant Quantum Enhanced Sensing and Imaging (QuEnSI), and by the National Research Council (NRC) of Canada via its Quantum Sensing Challenge Program.


%

\end{document}